# Plasmonic Enhancement of Second Harmonic Generation in Weyl Semimetal TaAs

Morris M. Yang[1,2], Mustafa Ozlu[1,2], Samuel Peana[1,2], Vahagn Mkhitaryan[1,2], Demid Sychev[1,2,3], Xiaohui Xu[2,4], Zachariah M. Martin[1,2], Hasitha Suriya Arachchige[5], Alexei Lagoutchev[2], David Mandurus[5], Vladimir Shalaev[1,2,3] and Alexandra Boltasseva[1,2,3,4]

[1]Elmore Family School of Electrical and Computer Engineering, Purdue University, West Lafayette, IN 47907, USA.
[2]Birck Nanotechnology Center, Purdue University, West Lafayette, IN 46907, USA.
[3]Department of Physics and Astronomy, Purdue University, West Lafayette, Indiana 47907, USA
[4]School of Materials Engineering, Purdue University, West Lafayette, IN 47907, USA.
[5]Department of Materials Science and Engineering, The University of Tennessee, Knoxville, TN 37996, USA
Email: aeb@purdue.edu

**Abstract:**

In this work a hybrid nanoplasmonic-Weyl Semimetal (WSM) structure is realized for the first time utilizing silver nanopatch antennas and WSM Tantalum Arsenide (TaAs). The studied hybrid WSM-nanoplasmonic structure demonstrated a substantial, over x4.5 enhancement of the second harmonic generation (SHG) process compared to a bare TaAs film. To realize the hybrid structure while preserving TaAs properties, a scalable, non-destructive manufacturing approach was developed that involves the fabrication of TaAs flakes from single crystalline TaAs, overgrowth of a silicon nitride overlayer, and drop-casting of silver nanopatch antennas. The strong polarization response of both the bare flakes, along with the hybrid-nanoplasmonic cavities demonstrates that this approach uniquely preserves the TaAs crystal structure and its optical response while providing significant enhancement of the nonlinear properties. The developed method allows leveraging the capabilities of plasmonics to control and enhance light-matter interactions at the nanometer scale to access and engineer WSM response. This work is the first step towards high-performance nanophotonic devices utilizing WSM topological properties.

*1. Introduction*

The Weyl equation describes the behavior of massless spin ½ particles called Weyl fermions [1], [2]. These particles are characterized by their linear dispersion and chirality. Despite intensive efforts to find elementary particles satisfying this criterion, none has yet been observed. However, in 1937, Conyers Herring, discovered that for materials with accidental band crossings, the dispersion relationship near a crossing point is well described by the Weyl equation [3]. In Weyl Semimetals (WSMs), a crossing point between the valence band and conduction band is known as a Weyl node, and the electronic quasiparticle corresponding to it has properties of a Weyl fermion [4], [5]. Due to the chirality of the Weyl node, such a solution in a solid material would violate charge conservation, thus in WSMs, Weyl nodes must always come in pairs of opposite chirality. Additionally, due to the nature of accidental band crossings, Weyl nodes are maintained even in the presence of perturbations. This resistance to perturbation can also be explained through lens of band topology by noting that Weyl nodes are monopoles of Berry curvature and are thus topologically protected. This rich electronic band structure manifests itself in a variety of interesting unusual phenomenon [6]–[10]. While there has been significant theoretical work studying the behaviour of WSMs, the first practical material exhibiting these properties, Tantalum Arsenide (TaAs), was only discovered in 2015 [11], [12]. This has led to an explosion of experimental work investigating the properties of bulk TaAs and other emerging WSMs. This has resulted in observation of many of the predicted WSM phenomenon including the confirmation of the giant photogalvanic effect[13], giant second harmonic generation [6], non-reciprocal flow of light, and other theoretically predicted effects[14]–[17].

Nanophotonics utilizing plasmonic structures offer unique ways to control and dramatically enhance light-matter interactions. Integration of such nanophotonic/nanoplasmonic structures to leverage and enhance the intrinsic properties of WSMs is emerging as a promising path to explore both new physics and novel device concepts. In this work, we for the first time demonstrate the integration of single-crystalline TaAs with a hybrid-plasmonic nanopatch antenna leading to a 466% enhancement in second harmonic generation (SHG) from the TaAs. Given that SHG is a second order nonlinear effect that scales with the amplitude of the electric field squared $|E(\omega)|^2$ the localization of light in a plasmonic nanostructure can result in enormous local SHG enhancement. Fundamentally, this is because surface plasmons with their extremely short wavelengths allow for the concentration of light far below the diffraction limit [18]–[28]. Moreover, since TaAs has a high refractive index (>3) at both the pump frequency (800nm) and the second harmonic wavelength of 400 nm [29] both the in-coupling and out-coupling efficiency of unstructured planar TaAs is poor. This is where plasmonic nanostructures can be designed to boost the coupling efficiency and the overall light extraction from the device [30]. The hybrid-plasmonic cavities on WSM TaAs demonstrated in this work greatly enhance the in-coupling of light and thus second harmonic generation.

This study also required the development of a non-destructive method of integrating nanoplasmonic structures with TaAs bulk crystals. Fabrication challenges associated with working with WSM are a major roadblock faced in realizing hybrid WSM-nanophotonic structures. Firstly, WSMs usually come in large, mm-sized crystals which are not compatible with standard planar nanofabrication approaches, such as electron beam lithography, reactive ion etching, etc. While WSM thin film growth is under development by various research groups, the existing growth of such materials results in oddly shaped TaAs pebbles that are not compatible with conventional processing equipment. An alternative approach we attempted was to use focused ion milling (FIB) to mill the bulk crystals directly. FIB is capable of nanostructuring materials without requiring them to be planar. However, we found that the properties of the TaAs degrade dramatically during such processes due to damage to the crystalline structure and undesirable doping with gallium (See Supplementary Section). In this work, we developed an non-destructive, large-scale approach of integrating plasmonic nanostructures with a TaAs WSM crystal by simply spin casting silver nanopatch antennas onto a carefully extracted flake of TaAs. We found that this method introduces no detrimental changes to the WSM properties.

## 2. Results and Discussion

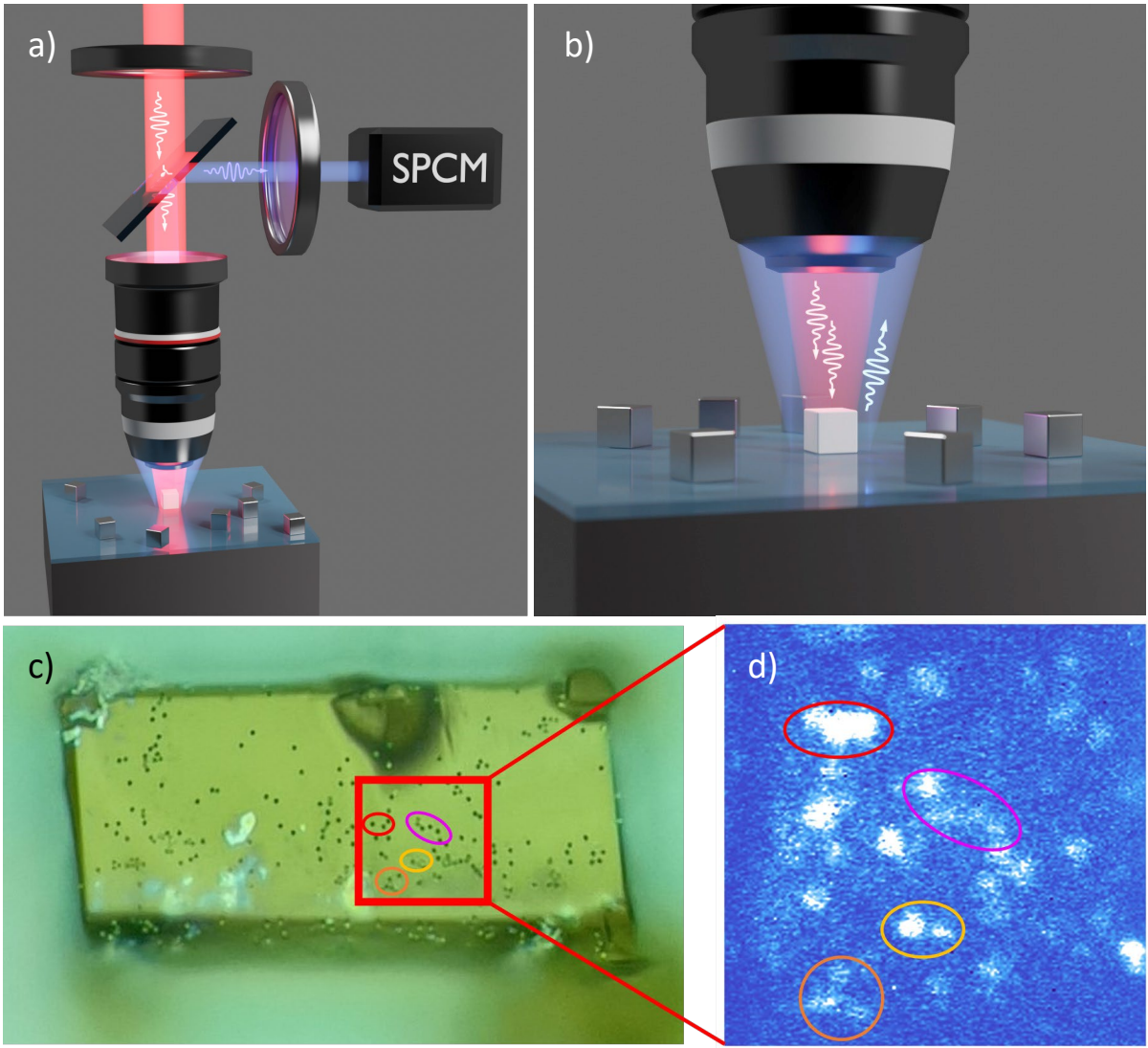

*Figure 1.* **Experimental setup. a.** An overview of the experimental setup, femtosecond laser pulses are sent through a half-wave plate, and focused by a 100X high NA objective onto a single silver cube dispersed on a TaAs surface. The generated SHG signal is then collected by the same objective, and directed to into a single photon counting module (SPCM) using a dichroic mirror. The pump signal is filtered out directly before the SPCM module. **b)** A close-up view of the silver nanocubes, on the SiN coated TaAs, the blue layer is the 25-nm-thick SiN, and the grey layer below is the TaAs. Note that the pump is focused on and the collected signal come from only one nanocube. **c)** An optical microscopy image of a TaAs flake coated with 25-nm-thick layer of SiN with silver cubes spin-casted on top. The black dots are silver cubes. **d)** A spatial map of the SHG signal intensity. The colors of the circles highlight the correspondence between the enhanced spots in the SHG map and the presence of nano cubes in the optical microscopy image.

TaAs is grown using high temperature chemical vapor transport process in the form of millimeter sized single-crystal "pebbles" with facets corresponding to the crystal faces. As noted above, both fabrication and experimentation on such non-planar substrates is exceedingly challenging. To circumvent these issues, a Gallium focused ion beam (FIB) is used to cut out 30-um-thick flakes of TaAs which are then transferred to a silicon substrate for easy handling [31]. Critically, this method of flake fabrication does not directly expose the face of the TaAs flake to the ion beam, to avoid the inevitable damage to the crystalline structure and Gallium contamination. This procedure results in TaAs flakes mounted onto a silicon substrate which can be used in standard nanofabrication equipment. This includes the high-density plasma enhanced chemical vapor deposition (HDPCVD) machine used to grow the 25 nm thick dielectric SiN spacer layer on the TaAs flake. Finally, 100nm silver nanocubes are dropcasted onto the sample to form nano-plasmonic cavities (NPCs). An optical microscopy image of the resulting system can be seen in Figure 1c. The black dots are the cubes and the rectangular piece they are on is the SiN coated TaAs flake.

The flake is created such that the <112> facet of the crystal is facing up. The SHG is spatially mapped using a homebuilt raster scanning confocal microscope (See supplement for more details). The enhancement of the SHG by the NPCs is clearly visible against the background SHG signal as can be seen in Figure 1c and d. Once a suitably isolated nanocube is found detailed SHG measurements are performed. An artistic rendering of the experimental setup can be seen in Figure 1a&b.

## 2.1 SHG Generation Enhancement

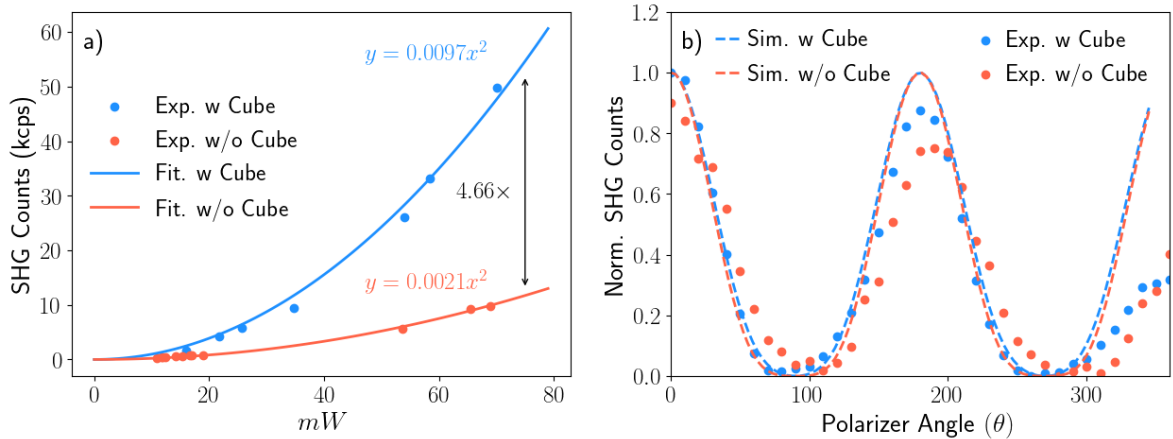

*Figure 2* **Experimental results. Left:** SHG counts rate dependence on the pump power. **Right:** The normalized SHG count rate dependence on the polarization of the pump at a fixed power level. The red dots correspond to the experimentally measured SHG count rate dependence for the bare SiN coated TaAs, the red dotted line corresponds to the simulated SHG count rate dependence of the same bare SiN coated TaAs. The blue dots are for the measured dependence for the nano-cube enhanced SHG. The blue dotted line corresponds to the simulated dependence for the same nano-cube enhanced structure.

The SHG pump power dependence was measured at the polarization where the SHG signal is at a maximum ($\phi = 0°$). The measured dependence is quadratic as expected since for SHG, $P(2\omega) \propto \chi^{(2)}_{eff}(\phi = 0)P(\omega)^2$. Where $\chi^{(2)}_{eff}(\phi = 0°)$ is the contribution of the different components of the anisotropic nonlinear susceptibility tensor of TaAs. This is true for both the SiN coated. TaAs surface and for those enhanced by NPCs as can be seen in Figure 2a. The SHG count rate is enhanced by 4.66X due to the NPC. This was the best enhancement factor we found during the course of our study with other enhancements being around 2-3X (see Supplementary section). This dramatic enhancement can be explained by examining Finite Difference Time Domain simulations (FDTD) simulations of this system using linear and nonlinear susceptibility tensors from literature. The simulations replicate key features of the experimental results closely enough to give insight into the mechanism of enhancement. In this simulation the input polarization of the pump is at 45° relative to the x-axis, this corresponds to the polarization with the maximum signal ($\phi = 0°$). By looking at the electric field intensity in the XY plane (Figure 3a) 1nm into the TaAs (Z-axis is normal to the TaAs <112> plane) there is strong enhancement of the penetration of the pump field into the TaAs under the nano-cube. This is particularly evident when looking at the XZ plane (Figure 3b), which clearly shows a deeply subwavelength gap plasmon mode occurring beneath the nano-cube. When compared with the electric field penetration into the TaAs without the cube (Figure 3c) it is apparent that the NPC dramatically increases in-coupling and concentration of the pump field into the TaAs.

Since the SHG is dependent on the electric field intensity the concentration of the pump electric field beneath the NPC leads to a significant increase in the SHG. This is evident from Figure 3c where the SHG electric field is plotted 1nm into the TaAs. The generated SHG spatial distribution matches the spatial distribution of the pump electric field as expected and is concentrated beneath the nano-cube. The cross section of the same system along the XZ plane (Figure 3b) shows that the SHG signal strongly outcouples into the gap between the nano-cube and the TaAs. However, looking at the field intensity above the cube it is apparent that many of the photons generated are being absorbed either by the silver cube or by the TaAs before they can out-couple to the far field to be detected. Despite this the number of SHG photons generated and emitted to the far field is significantly larger compared with the bare SiN coated TaAs (Figure 3e). In Figure 3e, the normalized electric field is multiplied by 10 otherwise the electric field intensity is not even visible since this plot shares the same color normalization with Figure 3c, d. From these simulations the expected enhancement in the SHG generation ~10X higher with the NPC compared to the case without the cube. This is in line with our experimental results where we observed a maximum enhancement ~4.66X due to the NPC. These variations in the enhancement factor are likely due to the dependence on the exact orientation of the cube relative to the crystalline structure of the TaAs, the exact size of the nanocube, and other variations in the sample.

## 2.2 SHG Polarization Dependence

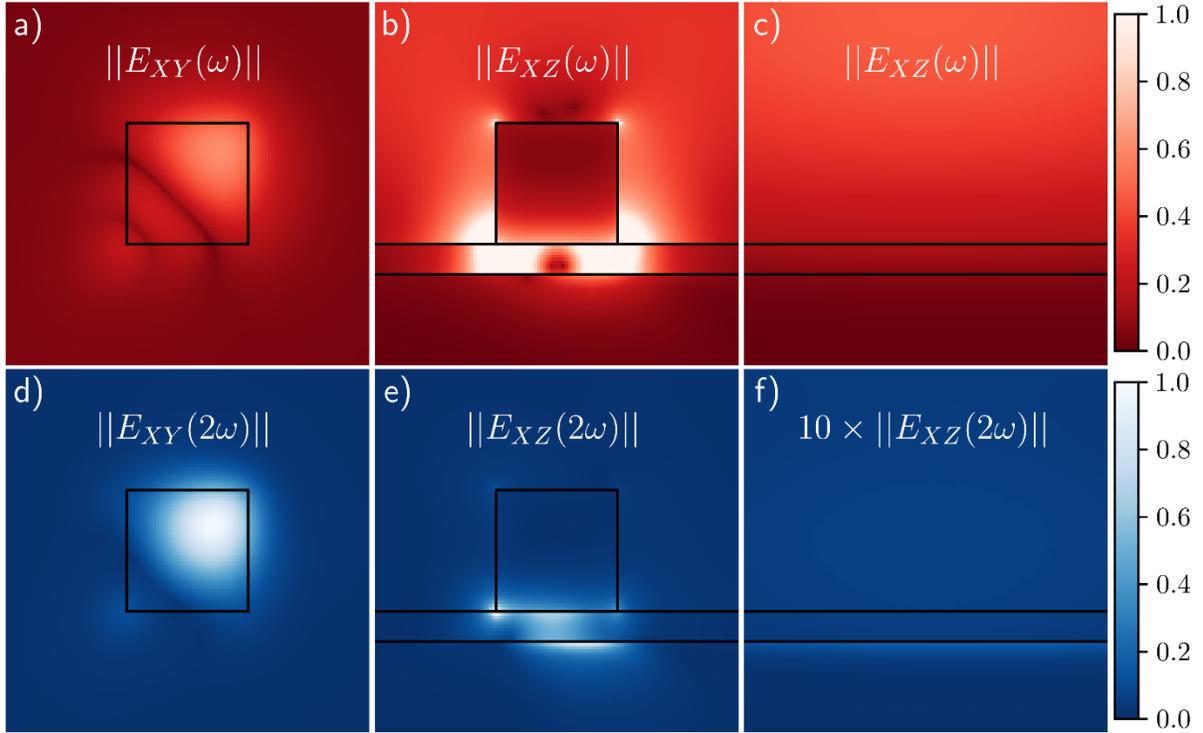

*Figure 3* **Simulation Results.** Plots of the normalized electric field at both the pump (top row) and second harmonic frequencies (bottom row). Top row, **a)** $|E(\omega)|$ 1nm into the TaAs in the XY plane and **b)** through the center of the nanocube in the XZ plane. **d)** $|E(\omega)|$ plotted for the case without the cube in the XZ plane. Bottom row, **d)** $|E(2\omega)|$, 1nm into the TaAs in the XY plane and **e)** through the XZ plane. **f)** $10 \times |E(2\omega)|$ plotted for the case without the cube in the XZ plane. The factor of 10 is to highlight the distribution of the electric field, otherwise it is too dim to see. The electric field amplitude distribution is identical for the XZ and YZ planes at both the pump and second harmonic frequencies. The cube in the figure is 100nm in width for scale. One thing to note, all of the field profiles in each row share the same color normalization so electric field amplitudes can be directly visually compared across the row.

TaAs is a highly anisotropic material in both its linear and nonlinear optical properties [32]. As a result, the SHG generated by TaAs has a strong dependence on the polarization of the pump field. To study this effect, a half-waveplate was used to rotate the pump polarization incident on the sample and the resulting changes in the SHG count rates were recorded. A typical result of this measurement is plotted in Figure 2b. The counts were normalized for both the bare film and the cube enhanced signal to highlight that both the bare film and nano-cube plasmonic cavity display almost the same polarization dependence. The most notable feature in both the simulation and experimental results is the strong modulation of the count rate with polarization. The good match between the normalized polarization dependence of the simulations, experimental results, and other literature of the SiN coated TaAs indicates conclusively that the planarized TaAs used in this work has maintained it strong anisotropy. Additionally, the silver cubes used in the NPC have four-fold symmetry and the amorphous SiN dielectric spacer has isotropic dispersion. Thus, these two components of the NPC will have no strong polarization dependent response of their own. This indicates that the polarization dependence of the NPC is almost entirely a result of the anisotropy of the TaAs in both its linear dispersion and second harmonic coefficients.

The strong polarization dependence of the SiN coated TaAs and the nanocavity enhanced SHG signal indicates that the method of fabricating TaAs flakes does not significantly degrade the crystalline structure of the TaAs as the anisotropic response of the TaAs is preserved. Additionally, the polarization dependence of the NPC also demonstrates that the fabrication processes involved in creating the NPCs also do not damage the TaAs crystalline structure either.

3. **Conclusion**

In this work, we for the first time successfully implemented a hybrid WSM-plasmonic structure that exhibits a substantial enhancement of the second harmonic generation process in WSM Tantalum Arsenide compared to bare TaAs. The developed hybrid structure involves a nanoplasmonic antenna non-destructively mounted on a dielectric spacer coated onto the TaAs surface, resulting in an effective nano-plasmonic gap cavity. In the developed planar, scalable fabrication method we first create TaAs flakes using Focused Ion Beam (FIB) undercutting, followed by over-growth of a Silicon Nitride (SiN) layer and drop-casting silver nanocubes. Importantly, this non-invasive technique does not degrade the TaAs crystal structure or its unique properties. Furthermore, this approach of generating thin (~30um) TaAs flakes allows for a single TaAs crystal to be used in common planar nanofabrication, paving the way for the future development of more complex hybrid nanophotonic devices.

We demonstrated the significant enhancement of the already inherently high SHG in TaAs by utilizing a single nanoplasmonic cavity. The SHG signal enhancement was found to be ~466% compared to a simple SiN coated TaAs film. We note that the nanoplasmonic cavity is formed by a silver nanocube measuring only 100 nm per side. Importantly, our experiments revealed that plasmonic structures, such as the studied nano-cavities, can probe the complex optical properties of WSM crystals. This is evidenced by the fact that the polarization response of the SiN coated TaAs and the NPC had the same polarization dependence. Effectively, the NPC served to dramatically amplify the existing anisotropic optical properties of the TaAs by allowing for more effective in-coupling of light into the high refractive index TaAs.

This work marks a critical step towards the development of a novel class of hybrid WSM-nanophotonic devices. Such devices will utilize both the distinctive features of nanoplasmonics that allows for unprecedented control and enhancement of light localization and the extraordinary inherent topological properties of WSMs.

4. **Methods**

*Crystal growth:* A Single crystal of TaAs was grown to mm-scale dimensions. Iodine was used as transport agent mixing with high purity tantalum and arsenic powders at temperature above 900°C to form polycrystalline TaAs. The ampule containing TaAs then went through chemical vapor transport at high temperature for periods up to 2 weeks before cooling into single crystal

In the CVT process, TaAs powder was synthesized by solid-state reaction at ~1000 °C for 48-72 hours after sealing mixed, ground, and pelletized stoichiometric quantities of Ta powder and pulverized As in a fused silica tube at a low-pressure Ar atmosphere. Care was taken when increasing temperature to avoid high partial pressure due to As. Then, after checking phase purity, pulverized TaAs and the transport agent $I_2$ were sealed in a 15 mm diameter fused silica ampoule with a length of ~11 cm at a low-pressure Ar atmosphere. The sealed ampoule was positioned horizontally in a two-zone furnace for single crystal growth such that the reagents were at the cold end and crystals were grown at the hot end of the ampoule. After placement of the ampoule, the temperature of the center of the two-zone furnace was gradually increased to ~1000 °C and dwelled for 7-14 days for crystal growth. The temperature gradient was about 50 °C. Finally, the ampoule was cooled to room temperature, and crystals with 0.5-2mm dimensions were separated [33], [34].

*TaAs Flake Preparation:* A crystal of TaAs was cut into a 60μm x 140μm, 30μm thick flake via $Ga^+$ focused ion beam (FIB). The flake was transferred to a Si substrate by Omniprobe. This method of cutting does not directly expose the working surface leaving it mostly unaffected by ion beam [35].

*Measurement of SHG*: A confocal microscopy setup with 100x and 0.9NA objective (Fig.1**b**) was used. Linearly polarized 100fs pulses from an 800nm laser (MaiTai) were used as the pump. The spot size was approximately 1um at the sample plane. The signal was collected by the same objective and measured with a single-photon detector (Excelitas, SPCM). The pump is filtered from the signal using a 550nm short-pass dichroic mirror and two filters (400/40nm bandpass and 550 short-pass). The objective is mounted onto a piezostage(PI-561.3CD), and is used for raster scanning the SHG signal and alignment to a particular cube for SHG measurements. The polarization dependence is probed using a motorized half-wave plate to rotate the pump polarization.

*FDTD Simulation*: The TaAs silver cube hybrid-plasmonic resonator system is simulated using the commercial FDTD software Lumerical. The input pump field was modelled as a polarized gaussian laser beam with a beam waist of 1um. The silver cube was simulated using the permittivity from [36]. The SiN was modelled using the permittivity from ellipsometry. The linear permittivity of the TaAs and the nonlinear susceptibility tensor used is found in [29].

**Acknowledgements:**

This work is supported by the AFSOR MURI Grant FA9550-20-1-0322s.